\documentclass[prl,twocolumn,showpacs]{revtex4}

\usepackage{graphicx}
\usepackage{amsmath}
\usepackage{epstopdf}
\usepackage{amsmath}
\usepackage{amsfonts}

\usepackage{epsfig}

\begin{document}

\title{Universal entanglement crossover of coupled quantum wires}

\author{Romain Vasseur$^{1,2}$, Jesper Lykke Jacobsen$^{3,4}$ and Hubert Saleur$^{5,6}$}

\affiliation{${}^1$ Department of Physics, University of California, Berkeley, Berkeley CA 94720, USA}
\affiliation{${}^2$ Materials Science Division, Lawrence Berkeley National Laboratory, Berkeley CA 94720, USA}
\affiliation{${}^3$LPTENS, 24 rue Lhomond, 75231 Paris, France}
\affiliation{${}^4$Universit\'e Pierre et Marie Curie, 4 place Jussieu, 75252 Paris, France}
\affiliation{${}^5$Institut de Physique Th\'eorique, CEA Saclay,
91191 Gif Sur Yvette, France}
\affiliation{${}^6$Department of Physics,
University of Southern California, Los Angeles, CA 90089-0484, USA}

\date{\today}

\begin{abstract}

We consider the entanglement between two one-dimensional quantum wires (Luttinger Liquids) coupled by tunneling through a quantum impurity.  The  physics of the system involves a crossover between weak and strong coupling regimes characterized by an energy scale $T_B$, and methods of conformal field theory therefore cannot be applied. The evolution of the entanglement in this crossover has led to many numerical studies, but has remained little understood, analytically or even qualitatively. We argue in this Letter that the correct universal scaling form of the entanglement entropy $S$  (for an arbitrary interval of length $L$ containing the impurity) is $\partial S/\partial \ln L = f(L T_B)$. In the special case where the coupling to the impurity can be refermionized, we show how the universal function  $f(L T_B)$ can be obtained analytically using recent results on form factors of twist fields and a defect massless-scattering formalism. Our results are carefully checked against numerical simulations.
  
\end{abstract}


\maketitle

\paragraph{Introduction.}
The study of two one dimensional gapless systems connected by some sort of interaction has become paradigmatic in modern quantum physics. It plays a particularly important role in the context of local quenches, transport through quantum dots, and the dynamics of magnetic impurities. 

An essential feature of these systems is the existence of crossover scales,  which play a role similar to  the Kondo temperature in the Kondo problem~\cite{Hewson}, and qualitatively  separate weak and strong coupling regimes. These scales make the methods of conformal field theory inapplicable, and exact results are very scarce. At the same time, the presence of the crossover indicates very rich physics.

A case in point is the so-called Kane-Fisher problem~\cite{KaneFisher}, where a single impurity in a Luttinger liquid has the dramatic effect of decoupling the two sides (repulsive case) or disappearing (attractive case) as energy  is swept across the crossover scale $T_B$. This problem appears in various other guises, in particular in experiments where a fractional quantum Hall fluid is pinched by a gate voltage~\cite{WenFQHE,resFQHE}. A quantity of crucial interest is then the entanglement entropy $S$ of a region (of length $L$) bounded by the impurity with the rest of the system. In the case where the impurity (defect) is marginal~\cite{Peschel10}, or for some classes of strongly disordered critical points~\cite{RefealMoore}, one can argue that $S \propto \ln L$. However, when the impurity is characterized by a crossover scale $T_B$, general arguments show that $S$ has a logarithmic behavior only in the low and high energy limits, with different pre-factors. The question of how $S$ interpolates between these -- both qualitatively and quantitatively -- has remained largely open up to now. An early study~\cite{Levine} attempted a perturbative approach, with results in disagreement with  numerics~\cite{DMRGXXZ}. The problem was revisited several times (see {\em e.g.}~\cite{AffleckLaflorencie2} for a review) before it was realized that, in fact, the entanglement in this problem is non-perturbative (at $T=0$)~\cite{SSV,LoicIR}. 
Similar questions arise in the  -- maybe even more interesting physically -- case where the  tunneling between the Luttinger liquids takes place through a resonant level (quantum dot)~\cite{EmeryKivelson}.

Unfortunately, non-perturbative approaches are few, especially for the entanglement, which is essentially a non-local quantity. Even when problems are in appearance ``free'', and involve a quadratic fermionic hamiltonian, the non-locality of $S$ makes analytical calculations  difficult, much like those involving observables which are non-local in terms of the fermions ({\em e.g.}, the spin in the Ising model). We report in this Letter the solution of this problem in such a ``free'' fermionic case, which we obtain by the combination  of a massless form-factors approach and a factorized scattering description which involves both reflection and transmission channels. We obtain results over the whole crossover, which are extremely well matched by numerical simulations. We also give the scaling form of the entanglement, which we argue generalizes to interacting situations.

\paragraph{XXZ spin chains and impurities.} We consider two semi-infinite spin-$\frac{1}{2}$ XXZ spin chains (spinless interacting fermions) in the gapless Luttinger Liquid (LL)
phase (with anisotropy $-1< \Delta \leq 1$) connected through either a weak link, or a quantum dot (two successive weak links). The Hamiltonian is
\begin{equation}
H=\sum_{i=-\infty}^{-2}\vec{S}_i \cdot \vec{S}_{i+1}
+\sum_{i=1}^\infty  \vec{S}_i \cdot \vec{S}_{i+1} + H_{\rm imp} \,,
\label{Hamiltonian}
\end{equation}
where $\vec{S}_i \cdot \vec{S}_{i+1} = S_i^x S_{i+1}^x + S_i^y S_{i+1}^y + \Delta S_i^z S_{i+1}^z$ is shorthand for the anisotropic interaction. The tunneling between the two
interacting wires is described by $H^{\rm wl}_{\rm imp} = J^\prime S^+_{-1} S^-_{1}+h.c.$ in the weak link case, or
$H^{\rm dot}_{\rm imp} = J^\prime (S^+_{-1} + S^+_{1}) S^-_{0}+h.c. $ in the dot case, with $S^\pm=S_x \pm i S_y$. We work at zero temperature so that the system is in a pure state.

\begin{figure}[t!]
\includegraphics[width=1.0\linewidth]{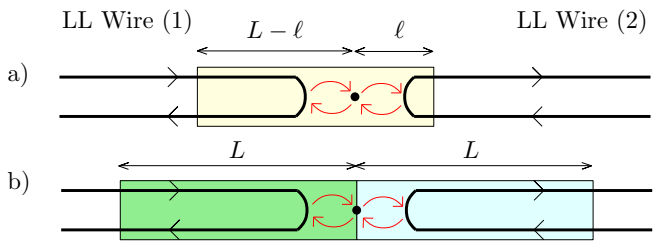}
\caption{Geometries considered in this Letter. We consider two Luttinger Liquids (LL) connected through an impurity, here a quantum dot. We are interested in:
 a) The Entanglement Entropy $S(\ell,L)$ of an interval, not necessarily centered on the impurity ($\ell \neq L/2$), and in particular in the limit $\ell \ll L$; and
 b) The Mutual Information $I(L)$ of two intervals of size $L$, with the impurity at their extremity. }
\label{fig1}
\end{figure}

\paragraph{Entanglement Entropy and Mutual Information.} To characterize the entanglement between the two wires, we consider two geometries
(Fig.~\ref{fig1}). We are mostly interested in the Entanglement Entropy (EE) $S(\ell,L)$ of an interval $\left[-L+\ell, \ell \right]$ of length $L$
not necessarily centered on the impurity (see Fig.~\ref{fig1}a). We characterize this asymmetry by the parameter $\alpha=\ell/L$. Recall that the EE can
be computed as $S= - {\rm tr} \rho \ln \rho$, where $\rho$ is the reduced density matrix obtained by tracing over the degrees of freedom outside
the considered interval. The symmetric case $\alpha=\frac{1}{2}$, which has been studied extensively
recently~\cite{AffleckLaflorencie1,AffleckLaflorencie2,EEimp1,EEimp2,EEimp3,LoicIR,SSV},
is natural for the Kondo problem, since the impurity lies at the boundary of a half-infinite chain in a folded picture \cite{EmeryKivelson}. However, the
limit $\alpha \to 0$ is more meaningful for two wires connected by an impurity, since it provides information on their entanglement. Another natural quantity is the Mutual Information (MI) $I(L)$ of two intervals of length $L$:
$A=\left[-L, 0 \right]$ and $B=\left[0, L \right]$ (see Fig.~\ref{fig1}b). The MI of $A$ and $B$, defined by
$I=S_A + S_B- S_{A \cup B}$, is positive and relates to the EE through $I(L) = 2 S(\ell=0,L) - S(\ell=L,2L)$.
The MI characterizes the correlations between two intervals, and provides an upper bound on their entanglement. In our problem, the MI
vanishes when the two wires are decoupled ($J'=0$). A full characterization of the entanglement between the wires would require
more precise estimators such as the negativity~\cite{Negdef} (see {\em e.g.}~\cite{Negativity1,Negativity2,Negativity3} in the context of the
Kondo problem), for which we expect the scaling predictions of this Letter to hold as well. We emphasize that the limit $\alpha \to 0$ of the EE is crucial when computing the MI, as it contains information on the entanglement between the two wires.

\paragraph{Bosonization and RG analysis.} We study the entanglement in the physically interesting case where the impurity is relevant
in the Renormalization Group (RG) sense. We consider energies much smaller than the band-width, where field-theoretic results are applicable.
The large-distance physics of the two XXZ half-chains can then be described by a LL with Luttinger parameter
$g^{-1}= 2-\frac{2}{\pi} \arccos \Delta$~\cite{LuttingerParameter}. After bosonization, the LL theory consists of a massless compactified boson, with right and left moving components scattering on the impurity. Unfolding the semi-infinite wires to obtain chiral bosons on the real line, one finds
\begin{equation}
H= \frac{v}{2 \pi}\int \sum_{a=1,2} {\rm d} x \, (\partial_x \phi_a)^2 + H_{\rm imp}[\phi_a(0)] \,,
\end{equation}
where $a=1,2$ labels the wires.
The impurity interaction reads $H^{\rm wl}_{\rm imp}= \lambda \cos \sqrt{2/g} \left( \phi_2(0)-\phi_1(0) \right) $
({\it resp.}  $H^{\rm dot}_{\rm imp}= \lambda  S_0^+ \sum_a \rm{e}^{-i \sqrt{2/g} \phi_a(0)} + h.c.$) in the weak link ({\it resp.} dot) case with  $\lambda \propto J^\prime+\dots$. Therefore~\cite{AffleckEggert} the weak link impurity has dimension $g^{-1}$ and is relevant for attractive interactions ($\Delta<0$, $g>1$) only. The system ``heals'' under renormalization, flowing to a strong-coupling fixed point
where the impurity is fully hybridized with the wires. The crossover is characterized by the energy scale $T_B \propto (J^\prime)^{1/(1-g)}$. Conversely, the dot impurity is always relevant, and at strong coupling the impurity is screened over a typical length scale $\xi_B \sim T_B^{-1}$ (the ``Kondo screening cloud''), with $T_B \propto (J^\prime)^{2/(2-g)}$.

Other impurity problems can be treated similarly, including the anisotropic Kondo problem, the interacting resonant level model, or the tunneling between Fractional Quantum Hall edges~\cite{WenFQHE,resFQHE,ConductanceFQHE}. The resulting chiral field theory can be folded back into an integrable boundary problem, which is usually convenient to perform calculations. In our case however, we stress that folding procedures are {\em incompatible} with the asymmetric geometry of Fig.~\ref{fig1}a, and one must maintain the original unfolded formulation, which is non-integrable in general.

\paragraph{Ultraviolet (UV) and infrared (IR) limits, perturbation theory.} The difficulty of computing the EE $S(\ell,L)$ in this impurity problem stems from
the energy scale $T_B$, and the asymmetric geometry. The weak and strong coupling limits can however easily be analyzed using Conformal Field Theory
(CFT) results~\cite{CardyCalabrese1,CardyCalabrese2}. In the weak coupling (UV) limit ($L,\ell \ll T_B^{-1}$) the physics is essentially given by two decoupled
wires with the interval at one boundary, so the EE reads
\begin{equation}
S_{\rm UV} \sim \frac{1}{6} \left[ \ln \frac{L-\ell}{a} + \ln \frac{\ell}{a}\right] \,,
\end{equation}
where $a$ is a UV cutoff (lattice spacing), and we have inserted the central charge $c=1$ of the LL liquid. For $\alpha=\frac{\ell}{L} \neq 0$, this
becomes $S_{\rm UV} \sim 2\times \frac{1}{6} \ln L$ for large $L$, whereas for $\alpha \to 0$ ({\it i.e.} $\ell \sim a$), one has instead
$S_{\rm UV} \sim \frac{1}{6} \ln L$, since the interval contains only a single half-wire. Clearly, $S_{\rm UV}$ contains non-universal terms when
$\alpha \to 0$, since the limiting procedure necessarily refers to the lattice spacing.

In the strong coupling (IR) limit ($L,\ell \gg T_B^{-1}$) the interval is in the bulk of a single healed wire, whence
\begin{equation}
S_{\rm IR} \sim \frac{1}{3} \ln \frac{L}{a} \,.
\end{equation}
Thus, for $\alpha \to 0$, the logarithmic term of the EE increases under
RG: $ \frac{1}{6} \ln L \longrightarrow \frac{1}{3} \ln L$. This increase is
expected, and witnesses to the ``healing'' of the chain upon renormalization.

Concerning the MI, one has $I_{\rm UV}=0$ at high energy, since the chains are decoupled, and
$I_{\rm IR} \sim \frac{1}{3} \ln L$ at low energy. Starting from these well-understood fixed points at tunneling amplitudes $\lambda=0$ and
$\lambda=\infty$, one could hope to compute the EE or the MI perturbatively.
However, the conclusions of~\cite{SSV}, obtained for a symmetric interval, apply to any $\alpha$.
The weak-coupling expansion of $S(\ell, L)$ is plagued by strong infrared divergences, indicating a non-analytic behavior in
$\lambda$, while the strong-coupling expansion can, in principle, be computed following~\cite{AffleckLaflorencie1, LoicIR},
although it would fail to capture the crossover physics. The non-perturbative nature of the weak-coupling expansion has unfortunately been overlooked previously.

\paragraph{Universal Scaling of the entanglement.}

Even though $S(\ell, L)$ (including the limit $\ell/L \to 0$) cannot be computed in general, one can still infer
its universal scaling form. For a symmetric interval ($\alpha=\frac{1}{2}$), it was argued~\cite{AffleckLaflorencie1} that $S(\ell =L/2,L)-S_{\rm IR}$ is a universal function of $L T_B$. Also in our case, we expect the EE to be related to a universal scaling function, interpolating between the weak and strong coupling regimes. However, it is clear from the evolution of the $\ln L$ term under the RG flow discussed above that $S(\ell,L)-S_{\rm IR}$ itself cannot be a scaling function of $L T_B$ for all values of $\ell/L$. Instead, we shall argue that the EE admits a general scaling
\begin{equation}
\frac{\partial S(\alpha=\ell/L,L)}{\partial \ln L} = f(L T_B, \ell/L),
\label{eqscaling}
\end{equation}
with $f(0, 0)=1/6$ and $f(0, \ell/L \neq 0)=1/3$ in the UV limit, and $f(\infty, \ell/L )=1/3$ at low energy. This scaling formula is physically appealing as it somehow follows the $\ln L$ term during the flow. Consequently $f(L T_B, \ell/L)$ can be thought of as some kind of ``effective central charge'', thus allowing a more precise interpretation of the numerics in~\cite{DMRGXXZ}, where a ``length-dependent effective central charge'' was introduced. One must be careful, however, since the derivative with respect to $\ln L$ obviously picks up other terms that are not logarithmic in $L$. 

Our main result~\eqref{eqscaling} can be obtained from the scaling of the Renyi entropy $S_n = \frac{1}{1-n} \ln R_n$, with $R_n = {\rm tr} \rho^n$ and $\rho$ the reduced density matrix introduced above. Recall that the EE can be computed from a replica trick as $S = - \left. \frac{\rm d}{{\rm d}n} R_n \right|_{n=1}$. The crucial point is the identification of $R_n$ as a two-point function of twist fields on a $n$-sheeted Riemann surface~\cite{CardyCalabrese1}. In our context of a $c=1$ CFT with a relevant boundary perturbation, we expect $R_n$ to scale as (dropping the $\ell$ dependence for simplicity)

\begin{equation}
R_n = {\rm tr} \rho^n = c_n \left( \frac{L}{a} \right)^{-\frac{1}{6}(n-n^{-1})} \Omega \left( L T_B, n \right) ,
\end{equation}
with $c_1 \Omega(L T_B,n=1)=1$ so that $R_1 = 1$. Here we have separated the universal scaling function $\Omega \left( L T_B, n \right)$ coming from the two-point function, and the non-universal proportionality coefficients $c_n $ that can be thought of as functions of $a T_B$ -- they can evolve during the flow, and they depend explicitly on  the UV cutoff $a$. The entanglement entropy can thus be expressed as $S =  h\left( L T_B\right) + k\left(a T_B \right)$ where $h\left( L T_B\right) =- \left. \partial_n \ln \Omega \right|_{n=1}+\frac{1}{3} \ln L T_B$ and $k \left(a T_B\right) = -\left. \partial_n \ln c_n \right|_{n=1}-\frac{1}{3} \ln a T_B$. To get rid of the non-universal contribution in the general case, we consider a derivative with respect to $\ln L$ to find~\eqref{eqscaling} as claimed.

\paragraph{Free Fermions Exact Solution.}

\begin{figure*}
\centering
\begin{tabular}{cc}
\epsfig{file=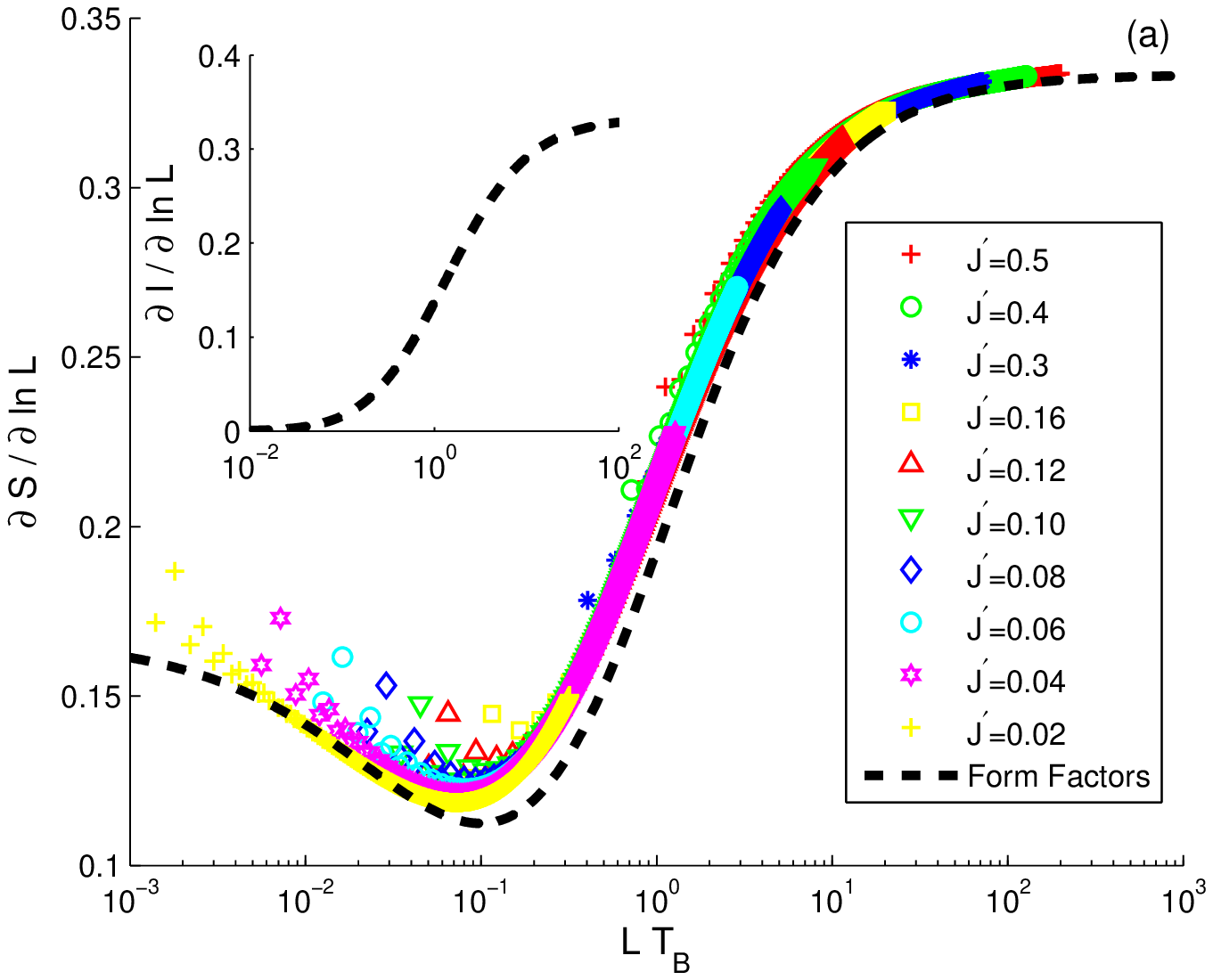,width=0.5\linewidth,clip=} &
\epsfig{file=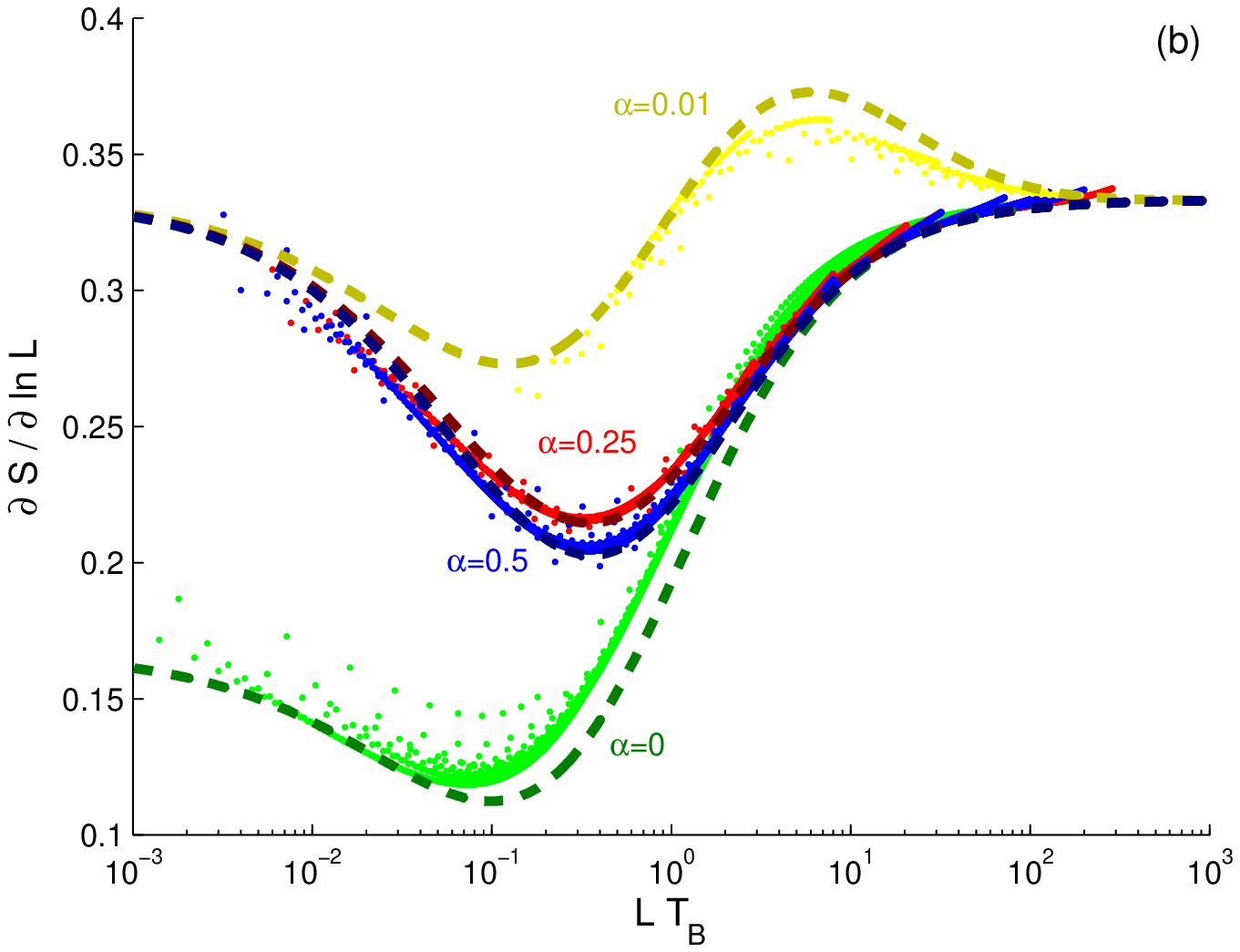,width=0.5\linewidth,clip=} 
\end{tabular}
\caption{EE scaling function $f(x,\alpha)$ with $x = L T_B$ and $\alpha = \ell/L$. The FF approximation
(\ref{MainResult}), shown as dashed lines, is compared with numerics for two wires of each $N=32\,000$ sites and
several values of $J'$.
(a) The limit $\alpha \to 0$. The inset shows the MI, without numerics.
(b) Different values of $\alpha$. For clarity, the numerical data for all values of $J'$ here carry the same color.}
\label{FigResults}
\end{figure*}

At the free fermion point $\Delta=0$, the scaling function~\eqref{eqscaling} can be computed exactly by combining the form factor approach of~\cite{Doyon1,Doyon} in the massless limit~\cite{SSV} with the defect scattering formalism~\cite{DMS}, where free particles are reflected and transmitted by the impurity with respective amplitudes $\hat{R}(\omega)$ and $\hat{T}(\omega)$ depending on their energy $\omega$. At leading order, one finds (see supplementary material)
\begin{multline}
S(\ell,L) \simeq - \frac{1}{4} \int_{0}^{\infty} \frac{{\rm d} \omega}{\omega} \mathrm{e}^{-2 L \omega} \hat{T}(\omega)^2 \\
 -\frac{1}{8} \int_{0}^{\infty} \frac{{\rm d} \omega}{\omega} \left[ \mathrm{e}^{-4 \ell \omega} +
 \mathrm{e}^{-4 (L-\ell) \omega} \right] \hat{R}(\omega)^2 \,. \label{DefectGeneral}
\end{multline}
The IR divergence of (\ref{DefectGeneral}) in the $\omega \ll 1$ limit can be cured by $\Gamma$-function
regularization techniques, or, more elegantly, by computing the logarithmic derivative (\ref{eqscaling}).

In the weak link case, one has $\hat{T}(\omega)^2 = \cos^2 \xi$ and $\hat{R}(\omega)^2=\sin^2 \xi$.
The parameter $\xi = \frac{\pi}{2} - 2 \arctan J'$ is independent of $\omega$, since the perturbation
is exactly marginal. In this case, $S(\ell,L)$ can however be computed exactly~\cite{Peschel10,Calabrese12},
so we turn instead to the more challenging dot case. (Note that the same formalism applies also to the weak link case with $\Delta=-\frac{1}{\sqrt{2}}$, which is interacting on the lattice but can be refermionized in the scaling limit). For the dot case at $\Delta=0$, one has
$\hat{T}(\omega)^2 = ( T_B / (T_B + \omega))^2$ and $\hat{R}(\omega)^2 = ( \omega / (T_B + \omega))^2$,
and we stress that $\hat{R}^2+\hat{T}^2 \neq 1$ only because unitarity has been broken by a Wick rotation in the computation leading to (\ref{DefectGeneral}). Our main result is then the lowest-order FF approximation to the
EE scaling function (\ref{eqscaling}):
\begin{multline}
f(x, \alpha) = \frac{2}{3} \int_{0}^{\infty} {\rm d} v \, \mathrm{e}^{- 2 v} \left( \frac{x}{x + v}\right)^2 \\
+ \frac{2}{3} \int_{0}^{\infty} {\rm d} v \, \left( \frac{\alpha}{\mathrm{e}^{4 \alpha v}} +
 \frac{1-\alpha}{\mathrm{e}^{4 (1-\alpha) v}} \right) \left( \frac{v}{x + v}\right)^2 \,, \label{MainResult}
\end{multline}
with the scaling variables $x = L T_B$ and $\alpha = \ell / L$. Notice that we have multiplied the actual result of
the computation by a factor $4/3$ in order to obtain the correct UV and IR limits~\cite{SSV}. This renormalization is justified {\em e.g.} by noticing that resummation of the full FF expansion in the UV/IR reproduces~\cite{Doyon,SSV} the known CFT result~\cite{CardyCalabrese1,CardyCalabrese2}. We note that the high-energy  $x \ll 1$ expansion of (\ref{MainResult}) contains an $x \ln x$ term for all $\alpha$, thus illustrating the non-perturbative nature of the EE, as already noticed for $\alpha=1/2$ in~\cite{SSV}. 

{F}rom (\ref{MainResult}) we also obtain the FF approximation to the MI scaling function:
\begin{multline}
\frac{\partial I(L)}{\partial \ln L} \equiv g(x)= \frac{4}{3} \int_{0}^{\infty} {\rm d} v 
\left( \frac{\mathrm{e}^{2 v} - 1}{\mathrm{e}^{4 v}} \right) \left( \frac{x}{v+x}\right)^2.
\end{multline}
In general the MI is only an upper bound on the entanglement between the two wires,
but in the limits $g(0) = 0$ and $g(\infty) = 1/3$ the bound is seen to saturate.

\paragraph{Numerical results.}

The EE scaling function $f(x,\alpha)$ exhibits a rich, non-monotonic behavior in both variables
(see Fig.~\ref{FigResults}), with an especially singular -- and physically interesting -- limit $\alpha \to 0$. We now check the accuracy of the FF approximation (\ref{MainResult}) against extensive numerics on the XX spin chain ($\Delta = 0$) with two weak links. Mapping the problem onto free fermions~\cite{SML}, the reduced density matrix can be obtained by diagonalizing the correlation matrix $\left< c^\dagger_n c_m \right>$~\cite{Peschel}, which in turn can be computed exactly from one-particle eigenstates (see supplementary material). To avoid numerical instabilities, we used both double and $50$-digit numerical precision. Our largest computations, with two wires of $N=32\,000$ sites each, are shown in Fig.~\ref{FigResults}. To avoid boundary effects, we considered intervals of length $L < N/10$. The values of $S$ showed strong parity effects in $L$ which were attenuated by averaging data for $L$ and $L+1$.

\paragraph{Discussion.} 

The agreement between the FF approximation (\ref{MainResult}) to $f(x,\alpha)$ and the numerics is excellent,
extending to more than five decades in $x=L T_B$ and all values of $\alpha$, including the $\alpha \to 0$ limit. Note that our results agree without any free parameter, as the scale $T_B = (J')^2$ can be computed exactly for $\Delta=0$. 
The considerable qualitative differences between $\alpha = 1/100$ and $\alpha = 0$ are well reproduced by the
numerics. (The case $\alpha=0$ is realized numerically by letting the interval start at the quantum dot site.) The scaling collapse for different values of $J'$ is remarkable, except for very small $x$ (high energy) where the lattice discretization is manifest. Presumably the small remnant discrepancies with (\ref{MainResult}) would disappear by taking the FF computation to the next order (see~\cite{SSV} in the $\alpha=1/2$ case).

More importantly, our universal scaling prediction~\eqref{eqscaling} goes beyond free-fermion systems -- or interacting systems that can be mapped onto free-fermions at low energy, and provides the correct description of entanglement in quantum impurity systems characterized by a Kondo temperature $T_B$. It would be very interesting to generalize this prediction to non-equilibrium setups, for example in the context of quantum quenches~\cite{LetterRLM}.

\smallskip

\paragraph{Acknowledgments.}
 This work was supported by the French Agence Nationale pour la Recherche (ANR Projet 2010 Blanc SIMI 4 : DIME), the US Department of Energy (grant number DE-FG03-01ER45908), the Quantum Materials program of LBNL (RV) and the Institut Universitaire de France (JLJ). We thank I.~Affleck, E.~Boulat, B.~Doyon, J.~Dubail, L.~Freton and P.~Schmitteckert for discussions.

\newpage

\centerline{\bf Supplementary material}

\bigskip

\section{Massless form factors formalism}

We describe here the Form Factor (FF) formalism leading to the computation of the entanglement entropy in the non-interacting case. The EE can be computed from a replica trick as $S = - \left. \frac{\rm d}{{\rm d}n} {\rm tr} \rho^n \right|_{n=1}$.
The replica limit $n \to 1$ is taken by analytic continuation from $n \in {\mathbb N}$, where the object ${\rm tr} \rho^n$
can be realized on an $n$-sheeted Riemann surface as a two-point function of branch-point twist fields $\tau_n$, $\tilde{\tau}_n$~\cite{CardyCalabrese1}.
The latter delimit the distinguished interval, along which sheets $i$ and $i+1$ are connected cyclically. Thus
$
S(\ell,L) \sim - \left. \frac{{\rm d}}{{\rm d}n}\langle \tau_n(-L+\ell)\tilde{\tau}_n(\ell) \rangle \right|_{n=1} \,.
\label{TwistReplica}
$
The computation of correlation functions of twist fields is in general extremely complicated, but for massive integrable models, a FF approach was developed in~\cite{Doyon1} for the bulk case, and specialized to the boundary Ising problem in~\cite{Doyon}. This framework provides a promising route to tackling our EE computation at
the free fermion point $\Delta=0$. Indeed, in the symmetric case ($\alpha=\frac12$)
folding the dot problem produces an Ising model with a boundary magnetic field, so that $S(\ell,L)$ reduces
to computing a one-point function~\cite{Doyon}, and the defect constitutes a known boundary state~\cite{GZ}.
The EE then follows by combining this and taking the massless limit~\cite{SSV}.
Folding around the impurity is incompatible with the general geometry ($\alpha \neq \frac12$), whence
these ingredients must be combined with the defect scattering formalism~\cite{DMS}, where free particles are reflected and transmitted by the impurity with respective amplitudes $\hat{R}(\theta)$ and $\hat{T}(\theta)$.

Using the notations of~\cite{Doyon1,Doyon}, the $k$-particle FF of a local operator ${\cal O}$ is denoted $F_k^{{\cal O}|\mu_1 \ldots \mu_k}(\theta_1,\ldots,\theta_k) = \left \langle 0 | {\cal O}(0) | \theta_1,\ldots,\theta_k \right \rangle^{\rm in}_{\mu_1,\ldots,\mu_k}$. Here $\mu_i$ refer to the Riemann sheets $i=1,\ldots,n$, and $\theta_i$ are the rapidities of the free fermionic particles. For the
twist field, ${\cal O} = \tau$, the replicated $S$-matrix is $S_{ij}(\theta) = \left( S(\theta) \right)^{\delta_{ij}}$, with $S(\theta)=-1$ in our case. The FF then satisfy
the fundamental relations~\cite{Doyon1}
$F_k^{\tau | \ldots \mu_i \mu_{i+1} \ldots}(\ldots,\theta_i,\theta_{i+1},\ldots) =
 S_{\mu_i,\mu_{i+1}}(\theta_{i,i+1}) F_k^{\tau | \ldots \mu_{i+1}\mu_i \ldots}(\ldots,\theta_{i+1},\theta_i,\ldots)$,
where $\theta_{ij} = \theta_i - \theta_j$, and
$F_k^{\tau | \mu_1\mu_2\ldots\mu_k}(\theta_1+2\pi i,\ldots,\theta_k) =
 F_k^{\tau | \mu_2,\ldots,\mu_n,\mu_1+1}(\theta_2,\ldots,\theta_k,\theta_1)$,
along with further axioms for the kinematic residue equations. This fixes the two-particle FF $F_2^{\tau | ij}(\theta_{12},n)$
as~\cite{Doyon1}
$F_2^{\tau | 11} =
-i {\langle \tau \rangle} \frac{\cos \left( \frac{\pi}{2n} \right) \sinh \left( \frac{\theta}{2n} \right)}{n \sinh \left( \frac{i \pi+\theta}{2n} \right)
 \sinh \left( \frac{i \pi - \theta}{2 n} \right)}$.
 
Neglecting for the moment the defect, one can then express
$\langle \tau_n(-L+\ell) \tilde{\tau}_n(\ell) \rangle$ in terms of FF by inserting a complete set of states at the position
of both twist fields. Truncating {\em e.g.} to two particles yields~\cite{Doyon1}
\begin{multline}
 \langle \tau_n(-L+\ell) \tilde{\tau}_n(\ell) \rangle \approx
 \langle \tau \rangle^2 + \frac{1}{2!} \sum_{i,j=1}^n \\ \int_{-\infty}^\infty \frac{{\rm d}\theta_1}{2 \pi} \frac{{\rm d}\theta_2}{2 \pi}
 \left| F_2^{\tau | ij}(\theta_{12},n) \right|^2 {\rm e}^{-L m(\cosh \theta_1 + \cosh \theta_2)} \,. \label{2partFF}
\end{multline}
The massless limit, $m \to 0$, is attained by setting $\frac{m}{2} = M {\rm e}^{-\theta_0}$ with $\theta_0 \to \infty$ \cite{SSV}.
Finite-energy excitations have $\theta = \pm(\theta_0 + \beta)$ with $\beta$ finite. 
They are L and R movers with momentum $p = \pm M {\rm e}^\beta$ and energy $e = |p|$.
Crucially, $F_2^{\tau | ii}(\theta,-\theta) \to 0$ for $m \to 0$, implying that each interval must sustain
an even number of L and R movers; this is related to the ${\mathbb Z}_2$ symmetry of the Ising model.

With the defect, these ingredients must be combined with the defect scattering formalism~\cite{DMS},
where free particles are reflected and transmitted by the impurity with respective amplitudes $\hat{R}(\theta)$ and $\hat{T}(\theta)$.
The lowest-order approximation (\ref{2partFF}) is then replaced by a sum of two diagrams. In the first,
two particles (RR or LL) are created at $\tau_n$, transmitted by the defect, and absorbed at $\tilde{\tau}_n$.
The corresponding contribution to $\langle \tau_n(-L+\ell) \tilde{\tau}_n(\ell) \rangle$ reads
\begin{multline}
\frac12 \sum_{i,j=1}^n \int_{-\infty}^{\infty} \frac{{\rm d}\theta_1}{2\pi} \frac{{\rm d}\theta_2}{2 \pi} \hat{T}(\theta_1) \hat{T}(\theta_2) \\
 \times \left| F_2^{\tau | ij}(\theta_{12},n) \right|^2 {\rm e}^{-L m (\cosh \theta_1 + \cosh \theta_2)} \,.
\end{multline}
In the second diagram, two LR pairs are created at $\tilde{\tau}_n(\ell)$ and reflected on the impurity, leading to
\begin{multline}
\langle \tau \rangle \sum_{i,j=1}^n \int \frac{{\rm d}\theta_1}{4\pi} \frac{{\rm d}\theta_2}{4 \pi} \hat{R}(\theta_1) \hat{R}(\theta_2) \\
 \times F_4^{\tau | iijj}(\theta_1,-\theta_1,\theta_2,-\theta_2,n) \, {\rm e}^{-2 \ell m (\cosh \theta_1 + \cosh \theta_2)} \,.
\end{multline}
The analogous process at $\tau_n(-L+\ell)$  is obtained by formally substituting $\ell \to L-\ell$. The $F_4^{\tau}$ FF can then be expressed in terms of $F_2^{\tau}$ using Wick's theorem~\cite{Doyon}.

Combining these diagrams one finds, in the massless limit,
\begin{multline}
S(\ell,L)= - \frac{1}{4} \int_{0}^{\infty} \frac{{\rm d} \omega}{\omega} \mathrm{e}^{-2 L \omega} \hat{T}(\omega)^2 \\
 -\frac{1}{8} \int_{0}^{\infty} \frac{{\rm d} \omega}{\omega} \left[ \mathrm{e}^{-4 \ell \omega} +
 \mathrm{e}^{-4 (L-\ell) \omega} \right] \hat{R}(\omega)^2 \,. \label{DefectGeneralSM}
\end{multline}
which is the main result announced in the text.

\section{Numerical methods}

After a Jordan-Wigner transformation, the XX spin chain becomes an itinerant fermion model with Fermi velocity $v_F=1$~\cite{SML},
for free fermions obeying canonical anticommutation relations $\{ c^\dagger_m,c_n \} = \delta_{mn}$.
We focus on the dot case, for which the impurity term reads $H^{\rm imp}_{\rm dot} = - J' \left( c^\dagger_1 c_0 + c^\dagger_0 c_{-1} + c^\dagger_0 c_1 + c^\dagger_{-1} c_0 \right)$ with the dot at $j=0$.

The one-particle eigenstates read $\left| \Psi_k \right> = \sum_j \phi_k(j) \left| j \right>$ with
$\left| j \right> = c^\dagger_j \left| 0 \right>$. The wave functions $\phi_k(j)$ obviously
take the form of R and L moving plane waves, $A \omega^j + B \omega^{-j}$, for $j \neq 0,\pm 1$,
where we have set $\omega = {\rm e}^{i k}$.
The Sch\"odinger equation reads $H \left| \Psi_k \right> = \epsilon_k \left| \Psi_k \right>$,
with the dispersion relation $\epsilon_k = -(\omega+\omega^{-1})$.
Solving this for $j=0,\pm 1,\pm 2$
provides 5 relations between the 4 plane wave amplitudes and the 3 amplitudes $\phi_k(j=0,\pm 1)$.
Normalizing $\phi_k(j)$, and imposing the boundary conditions $\phi_k(\pm N) = 0$ for a system of $2N-1$ sites,
then results in one quantization condition,
$ \left( \omega^{2N} - 1 \right) \left( 1 + \omega^2 -  2 J'^2 \omega^2 - \omega^{2N} (1 -  2 J'^2 + \omega^2 ) \right) = 0$.
This has precisely $2N-1$ solutions that we normalize by $\sum_{nm} \phi^*_k(n) \phi_{k'}(m) = \delta_{k,k'}$.
The two-particle correlation matrix $C(n,m) = \left< c^\dagger_n c_m \right>$ then reads
$\left< c^\dagger_n c_m \right> = \sum_{k < k_F} \phi^*_k(n) \phi_k(m)$, where the Fermi level $k_F = \pi/2$.

The numerical computation of the EE now proceeds as follows~\cite{Peschel}. The reduced density matrix
$\rho \propto \exp \left( - \sum_{ij} \Xi(i,j) c^\dagger_i c_j \right)$ is related to the correlation
matrix by $\Xi = \log \left[ (1-C)/C \right]$, with the indices of both constrained to the distinguished interval.
The eigenvalues $\zeta_\ell$ of $C$ satisfy $0 < \zeta_\ell < 1$ and we have
$S = - {\rm Tr}\, \rho \log \rho = - \sum_\ell \left[ \zeta_\ell \log \zeta_\ell + (1-\zeta_\ell) \log(1-\zeta_\ell) \right]$.
The diagonalization of $C$ is an ill-conditioned problem, most of $\zeta_\ell$ being exponentially close
to $0$ or $1$, but since their contribution to $S$ is negligible it suffices to use standard diagonalization libraries and standard numerical precision.

\end{document}